%% file: Bouma2.tex
\documentclass[preprint,12pt]{elsarticle}

\usepackage{amssymb}
\usepackage{amsmath}
\usepackage{url}
\usepackage{nccbbb}
\usepackage{extpfeil}
\usepackage{arydshln}
\usepackage[colorlinks=true]{hyperref}
\usepackage[ruled,vlined]{algorithm2e}
\usepackage{graphicx}
\usepackage{epstopdf}
\usepackage{needspace}
\usepackage{soul}
\newcommand{\eg}{e\@.g\@.~}                     % e.g.
\newcommand{\ie}{i\@.e\@.~}                     % i.e.

\newcommand{\singlequote}[1]{`#1'}
\newcommand{\quotes}[1]{``#1''}
\newcommand{\citeb}[1]{\cite[]{#1}}
\newcommand{\itbf}[1]{\textit{\textbf{#1}}}

\newcommand{\BBSigma}{\textbf{$\Sigma$ }}
\newcommand{\BBSigmaStar}{\textbf{$\Sigma^*$ }}
\newcommand{\BBSigmaTwo}{\textbf{$\Sigma^2$ }}
\newcommand{\BBword}{\textbf{$w$ }}
\newcommand{\BBwordlen}{\textbf{$|w|$ }}
\newcommand{\BBLanguage}{\textbf{$L$ }}
\newcommand{\BBLanguageSize}{\textbf{$sz(L)$ }}
\newcommand{\BBepsilon}{\textbf{$\epsilon$ }}
\newcommand{\BBsurjective}{\xtwoheadrightarrow{}}
\newcommand{\reg}{\textsuperscript{\textregistered}}

\newtheorem{thm}{Theorem}

\newtheorem{pro}{Proposition}

\newdefinition{defn}{Definition}
\newproof{pf}{Proof}
\newdefinition{fact}{Fact}
\newdefinition{claim}{Claim}
\newproof{pfclaim}{Proof of Claim}
\newdefinition{example}{Example}

\journal{arXiv.org}

\begin{document}

\begin{frontmatter}



\title{Bouma2 - A Quasi-Stateless, Tunable Multiple String-Match Algorithm}


\author{Erez M. Buchnik\corref{cor1}}
\ead{Erez.Buchnik@gmail.com}
\cortext[cor1]{Corresponding author}

%% \address{P.O. Box 9345, \\
%% Zoran, Israel}

\begin{abstract}
The Bouma2 algorithm attempts to challenge the prevalent \quotes{stateful} exact string-match paradigms by suggesting a \quotes{quasi-stateless} approach. We claim that using state-machines to solve the multiple exact string-match problem introduces a hidden artificial constraint, namely the \itbf{Consume-Order Dependency}, which results in unnecessary overhead. Bouma2 is not restricted in this sense; we postulate that this allows memory-efficiency and improved performance versus its state-machine equivalents. The heart of the Bouma2 preprocessing problem is formulated as a weighted \itbf{Integer Linear Programming} problem, that can be tuned for memory footprint and performance optimization. Specifically, this allows Bouma2 to be input-sensitive, as tuning can be based on input characteristics. Evaluating Bouma2 against the Aho-Corasick variant of the popular Snort Intrusion Prevention System, we demonstrate double the throughput while using about 10\% of the memory.
\end{abstract}

\begin{keyword}
pattern match \sep hash function \sep integer linear programming \sep motif \sep deep packet inspection \sep Snort


\end{keyword}

\end{frontmatter}
\date{}

% \linenumbers

\input{b2-introduction}

\input{b2-related-work}

\input{b2-basic-definitions}

\input{b2-preprocessing}

\input{b2-match-process}

\input{b2-experimental-results}

\input{b2-conclusion}

\input{b2-future-work}



%% References


\bibliographystyle{elsarticle-num}




\bibliography{b2-cites}



% \end{thebibliography}

\input{b2-appendix}

\end{document}

%% file: b2-introduction.tex
\section{Introduction}\label{sec:Introduction}
Multiple exact string-match is a classical problem with a vast range of applications and solutions. This paper describes the Bouma2 algorithm, which takes a somewhat unorthodox approach to this problem. The most significant difference is that Bouma2 is \itbf{Consume-Order Agnostic}: the majority of existing algorithms restrict their match procedure to consume symbols in a predefined order (usually left-to-right or right-to-left). This restriction is by no means implied from the definition of the problem (see Definition~\ref{defn:The Multiple Exact String Match Problem}). Indeed, existing algorithms would require certain data-structures to be built for left-to-right searches, but these data-structures would usually be useless for right-to-left searches over the same input. With Bouma2, the same data-structures can be used for matching in any sequence. We believe that this difference is what makes Bouma2 so efficient in memory: many redundant scenarios in the \quotes{event-horizon} of the match-procedure do not require consideration in the Bouma2 data structures.

In order to explain Bouma2 more easily, we try to intuitively relate it with some cognitive models that describe human word recognition. The following sentence can be easily understood by a person with basic English reading skills:

\begin{quote}
\begin{center}
\textit{\textbf{"If you can raed tihs,  \\ tehn you are prbbolay not a sttae-mhciane."}}
\end{center}
\end{quote}

The spelling mistakes in the text are also immediately noticeable. It is claimed that the brain first identifies the contour, or \textit{Bouma Shape} \citeb{bouma1973visual, saenger2000space, citeulike:7227}, of each word, already associating it with its possible meaning, and only then the spelling mistakes are considered. Conversely, a state-machine capable of recognizing every correctly-spelled word through a single left-to-right pass over the text would simply yield a no-match result. An alternative exact-match algorithm that maintains \quotes{hints} for every word (\eg for the above sentence the first and last character in every word) would still return a no-match result for misspelled words, but would also be capable of reporting \quotes{false-positives} (which may roughly be associated with spelling mistakes). This would have to be at the expense of traversing parts of the text more than once, contrary to the one-pass behavior that is typical of state-machines.

Bouma2 follows similar concepts: it searches first for \itbf{Motifs}\footnote{The term \textit{Motif} was adopted from Computational 
Biology \citeb{Lawrence1993}, where \textit{Sequence Motifs} are used in DNA sequence analysis schemes.}, which are 2-symbol substrings of any one of the sought strings. For every motif occurrence in the input string, symbols around the motif location are examined to corroborate the match. Bouma2 maintains a mapping between the original pattern strings and their corresponding motifs. For every pattern string there are 2 mappings: one mapping to a motif at an even offset from the beginning of the pattern string, and one mapping to an odd-offset motif. Obviously, many pattern strings may be mapped to the same motif. This would require a resolving process when such a motif is located in the input. For efficient resolving we have developed a new data-structure termed \itbf{the Mangled Trie} (see Section~\ref{subsec:Mangled-Trie Construction}).

This double mapping scheme allows the match procedure to advance in 2-symbol strides, and still find matches at any given offset. This, and the fact that every motif match is handled separately, together allow Bouma2 to be \itbf{Consume-Order Agnostic}: as long as all the input is consumed, and all the motif matches are accounted for, there is no importance to the exact order in which the input traversal is performed. Indeed, the match process could be completely parallelized by accessing all non-overlapping 2-symbol sequences at once; this obviously hints at a very efficient hardware implementation of this algorithm.

The mapping of the set of strings to their corresponding motifs can be viewed as a hash-function. For large numbers of strings, there may be millions of different valid motif-sets and hash-functions. We optimize the selection of the motif-set by means of formulating the problem as an \itbf{Integer Linear Programming} problem and solving it using the \itbf{Branch-and-Cut} algorithm. Furthermore, weights can be applied to every potential motif in order to optimize memory, performance etc.

The remainder of this paper is organized as follows: Section~\ref{sec:Related Work} surveys related work, with emphasis on consume-order dependency; Section~\ref{sec:BasicDefinitions} defines basic concepts and notations; Section~\ref{sec:The Bouma2 Preprocessing Stage} presents the 3 parts of the Bouma2 preprocessing stage; Section~\ref{sec:The Match Process} describes the match process and explains its separation to \textit{Fast-Path} and \textit{Slow-Path}; Section~\ref{sec:Experimental Results} documents benchmarks done against the \textit{Snort\reg} IPS software; Section~\ref{sec:Conclusion} provides conclusions and Section~\ref{sec:Future Work} describes future work. In~\ref{sec:Mangled-Trie Construction Algorithm} we provide the \textit{Mangled-Trie} construction heuristics and in~\ref{sec:Match-Process Algorithm} the \textit{Fast-Path} and \textit{Slow-Path} match-time procedures.

%% file: b2-related-work.tex
\section{Related Work}\label{sec:Related Work}
Multiple exact string-match is a classical problem with applications in many fields, ranging from Information Security \citeb{Al-Saleh:2011:ART:1972441.1972454, norton2004optimizing, bremler2010compactdfa, udhayanlightweight} through Internet Services \citeb{scarpazza2008high, bremler2011space}, Bioinformatics \citeb{dandass2008accelerating, sanders2011proteogenomic} and others. The diversity of the different solutions to this problem is impressive. In this paper we informally describe the notion of \itbf{Consume-Order Dependency}, which we claim is characteristic of the majority of existing solutions to this problem. A \textit{Consume-Order Dependent} solution would typically include a data-structure preprocessed from the pattern strings and a match-time algorithm that would traverse this structure along with the input string, but would also implicitly dictate the order of traversal over the input string. A \textit{Consume-Order Agnostic} solution, on the other hand, would be free of the traversal order constraint, and there would exist several equally efficient algorithms, which would be able to utilize the same preprocessed data-structure in different traversal orders, and still arrive at the same correct match results. Note that in this paper we do not give a formal definition of \textit{Consume-Order Dependency}, nor do we prove the superiority of \textit{Consume-Order Agnostic} solutions versus other solutions - the notion is used primarily for classification. Nevertheless, we intuitively suggest that in general, solutions imposing a constraint that is immaterial of the original problem they aim to solve, may potentially be less efficient than solutions that are free of this constraint.

The de-facto industry standard for multiple exact string match is the Aho-Corasick algorithm \citeb{aho1975efficient}. This, and its family of derived algorithms \citeb{commentz1979string, Baeza-Yates:1992:NAT:135239.135243}, variants and optimizations \citeb{bremler2011space, lin2010accelerating}, are all inherently \textit{Consume-Order Dependent}: they assume traversal of the input in atomic steps following a well-defined order, usually with no option of re-examining an already traversed symbol.

The Wu-Manber algorithm \citeb{wu1994fast} extends the Boyer-Moore \citeb{boyer1977fast} single-pattern match algorithm to multiple patterns. It belongs to a separate family of skip-table algorithms, which preprocess one or more tables for indicating how many symbols can be skipped based on previously consumed symbols. This family also has many variations, improvements and applications \citeb{ke2006improved, sun2006improved, lassmann2005kalign, navarro2002flexible}. Of course, all of them again exhibit \textit{Consume-Order Dependency}: the skip-table is calculated assuming that skips are always in the same direction. A different skip-table would have to be built if consume-order is reversed.

String-hashing algorithms like Rabin-Karp \citeb{karp1987efficient} and Muth-Manber \citeb{muth1996approximate} treat strings as keys to a hash-function, and use the hash-value to find a subset of possibly matching strings, all sharing the same hash-value. Here, \textit{Consume-Order Dependency} is a necessity if rolling-hash functions are chosen (which is usually the case, for performance reasons). Although the input can be consumed in ANY order, and the hash-function can be calculated at any point in the input, only a left-to-right order or a right-to-left order allow a more efficient rolling-hash calculation. Nevertheless, Bouma2 has considerable affinity to this family of algorithms, as it also maps strings to \quotes{hash-values} and resolves collisions at match-time. The differences are:

\begin{enumerate}
\item The hash-function is tailor-made using optimization techniques
\item Hash-values are always 2-symbol substrings of the key string
\item Every key string is mapped \itbf{twice} to hash-values
\item Collision-resolving takes into account the already matched 2-symbol substring for relative offset information
\end{enumerate}

Bouma2 is also affiliated with filtering algorithms like Q-Grams \citeb{salmela2007multipattern} and Bloom-Filters \citeb{Bloom70space/timetrade-offs, dharmapurikar2003deep, Moraru:FastCacheTR159}, which may exhibit \textit{Consume-Order Independence} during the filtering phase. Traversing the input in non-overlapping 2-grams is very close to the Bouma2 \textit{Fast-Path} procedure (see Section~\ref{sec:The Match Process}). Nevertheless, we have not seen any documented post-filtering phase that is \textit{Consume-Order Agnostic}. 

It is also important to note that our approach attempts to be input-sensitive at the expense of losing generality. The Bloom-Filter approach, for example, assumes that choosing a uniformly-distributed hash and reducing false-positives for the random case (see \citeb{kirsch2006less}) would improve performance. Our claim is that optimizations that are aware of input characteristics will be more effective in real-life situations.

%% file: b2-basic-definitions.tex
\section{Basic Definitions}\label{sec:BasicDefinitions}
An alphabet \BBSigma is a nonempty set of symbols. A word over \BBSigma is a finite sequence of symbols of \BBSigma. The empty word is denoted by \BBepsilon and the length of a word \BBword is denoted by \BBwordlen. The set of words over \BBSigma is denoted by \BBSigmaStar. A language \BBLanguage is any subset of \BBSigmaStar. The set of all 2-symbol words is denoted by \BBSigmaTwo. The total length of all words in a language $(\sum_{w\in L}|w|)$ is denoted by \BBLanguageSize (refer to \citeb{hopcroft1979introduction} for more details).

%%%%%%%%%%%%%%%%%%%%%%%%%%%%%%%%%%%%%%%%%%%%
\begin{defn}[The Multiple Exact String Match Problem]\label{defn:The Multiple Exact String Match Problem}
	Given a language $L\subseteq \Sigma^*$ and a long word $W_I\in \Sigma^*$, find all occurrences of any of the words in \BBLanguage that are substrings of $W_I$.
\end{defn}
%%%%%%%%%%%%%%%%%%%%%%%%%%%%%%%%%%%%%%%%%%%%

\subsection{Traces and Motifs}
Given a language, Bouma2 first identifies the complete set of all 2-symbol substrings of all the words (termed \itbf{Traces}). Then, every word is mapped to two of its own 2-symbol substrings, or \itbf{Motifs}. There are 2 mappings per word, one at an even offset and one at an odd offset. The following describes the relationships between words, traces and motifs.

%%%%%%%%%%%%%%%%%%%%%%%%%%%%%%%%%%%%%%%%%%%%
\begin{defn}[Trace-Set]
	For a given language \BBLanguage, the set $T_L\subseteq \Sigma^2$ that satisfies:
	\begin{equation}
	  t\in T_L \iff \exists w\in L \ \ and\ \ \exists w_{p},w_{s}\in \BBSigmaStar:w=w_{p}{t}w_{s} \;  .
	\end{equation}
	is named the \itbf{Trace-Set} of \BBLanguage. Any $t\in T_L$ is named a \itbf{Trace}.
\end{defn}
%%%%%%%%%%%%%%%%%%%%%%%%%%%%%%%%%%%%%%%%%%%%

%%%%%%%%%%%%%%%%%%%%%%%%%%%%%%%%%%%%%%%%%%%%
\begin{defn}[The Trace Occurrence Function]
	The function $occ$ that satisfies:
	\begin{eqnarray}
		occ & : & L \times \Sigma^2 \times \bbbn \to \{0,1\} \;  .  \nonumber  \\
	  occ(w,t,l) = 1\ & \iff & w=w_{p}{t}w_{s} \land |w_{p}|=l \;  .
	\end{eqnarray}
	is named the \itbf{The Trace Occurrence Function} of \BBLanguage.
\end{defn}
%%%%%%%%%%%%%%%%%%%%%%%%%%%%%%%%%%%%%%%%%%%%

%%%%%%%%%%%%%%%%%%%%%%%%%%%%%%%%%%%%%%%%%%%%
\begin{defn}[The Trace Association Functions]
	The following functions, $assoc_0$ and $assoc_1$, are respectively named \itbf{the Even and Odd Trace Association Functions}, and are defined as:
	\begin{eqnarray}
		assoc_0,assoc_1 & : & L\times \Sigma^2 \to \{0,1\} \;  ,  \nonumber  \\
		assoc_0(w,t) = 1 & \iff & \sum_{l=0}^{\lfloor |w|/2 \rfloor}occ(w,t,2l) > 0 \;  .  \\
		assoc_1(w,t) = 1 & \iff & \sum_{l=0}^{\lfloor (|w|-1)/2 \rfloor}occ(w,t,2l+1) > 0 \;  .
	\end{eqnarray}
\end{defn}
%%%%%%%%%%%%%%%%%%%%%%%%%%%%%%%%%%%%%%%%%%%%

%%%%%%%%%%%%%%%%%%%%%%%%%%%%%%%%%%%%%%%%%%%%
\begin{defn}[Motif-Set]\label{defn:Motif-Set}
	Given a trace-set $T_{L}$, any $M_{L}\subseteq T_{L}$ that satisfies for every $w\in L$:
	\begin{eqnarray}
	  \sum_{t\in M_{L}}assoc_0(w,t) \geq 1 \land \sum_{t\in M_{L}}assoc_1(w,t) \geq 1 \;  .
	\end{eqnarray}
	is named a \itbf{Motif-Set} of \BBLanguage. A \itbf{Motif} is every trace $\mu\in M_L$.
\end{defn}
%%%%%%%%%%%%%%%%%%%%%%%%%%%%%%%%%%%%%%%%%%%%

%%%%%%%%%%%%%%%%%%%%%%%%%%%%%%%%%%%%%%%%%%%%
\begin{pro}[Motif-Set Existence]
	There exists a motif-set for every $L$ satisfying:
	\begin{equation}
	  \forall w\in L:|w|>2 \;  .
	\end{equation}
	\begin{pf}
		By example: consider the trace-set $M_{L}$, in which the condition in Definition~\ref{defn:Motif-Set} is inherently satisfied:
		\begin{eqnarray}
		  M_{L}:= \bigcup_{w\in L} \{t:occ(w,t,0)=1\} \ \cup \ \bigcup_{w\in L} \{t:occ(w,t,1)=1\} \;  .
		\end{eqnarray}
		\qed
	\end{pf}
\end{pro}
%%%%%%%%%%%%%%%%%%%%%%%%%%%%%%%%%%%%%%%%%%%%

\begin{example}[{{\rm Motif-Sets}}]\label{ex:Motif-Sets}
%%%%%%%%%%%%%%%%%%%%%%%%%%%%%%%%%%%%%
Consider the following 6-word language of 8-bit characters.
\begin{equation}
  L=\{ \verb|herd,herbal,upper,deeper,error,ferrarri|\footnote{Intentional misspelling.} \} \;  .
\end{equation}
Table~\ref{tbl:Motif-Sets} shows the trace-set for this language. It also shows 3 motif-sets: the first one attempts to map all the words using as few traces as possible, the second tries to avoid traces that are frequent in the word-set, and the third tries to avoid traces that are expected to be frequent in the input (using trace occurrence statistics collected in advance). See Section~\ref{subsec:Optimizing Motif Selection} for an explanation of the motif-selection process.

\begin{table}
%%%%%%%%%%%%%%%%%%%%%%%%%%%%%%%%%%%%%%%%%%
\begin{center} \small
\begin{tabular}{|c|c|c|c|}
\hline
$T_L$ & $M_L$ & $M_L$ & $M_L$ \\ 
  & (minimum motifs) & (prefer rare motifs & (prefer rare motifs \\ 
  &  & in strings) & in input) \\ 
\hline
& & & \\[-8pt]
\verb|fe| &  & \verb|fe| &  \\
\verb|ee| &  & \verb|ee| &  \\
\verb|ep| &  & &   \\
\verb|ra| &  & &   \\
\verb|rb| &  & \verb|rb| & \verb|rb| \\
\verb|rd| &  & \verb|rd| & \verb|rd| \\
\verb|ri| &  & &   \\
\verb|ro| &  & &   \\
\verb|al| &  & &   \\
\verb|ar| &  & &   \\
\verb|ba| &  & &   \\
\verb|up| &  & \verb|up| &  \\
\verb|pp| &  & &   \\
\verb|de| &  & &   \\
\verb|or| &  & \verb|or| &  \\
\verb|he| & \verb|he| & &   \\
\verb|pe| & \verb|pe| & & \verb|pe|  \\
\verb|rr| & \verb|rr| & & \verb|rr|  \\
\verb|er| & \verb|er| & \verb|er| & \verb|er| \\
\hline
\end{tabular}
\end{center}
\caption{\itbf{Motif-Sets:} The leftmost column shows the complete trace-set for Example~\ref{ex:Motif-Sets}, while the other columns specify 3 different selections of motifs. The first selection attempts to form the smallest possible motif-set. The second selection tries to find motifs that appear as few times as possible within the pattern strings. The last selection tries to find motifs that ar expected to appear as few times as possible within the expected input string (based on statistics collected in advance).}~\label{tbl:Motif-Sets}
%%%%%%%%%%%%%%%%%%%%%%%%%%%%%%%%%%%%%%%%%%
\end{table}

%\begin{table}
%%%%%%%%%%%%%%%%%%%%%%%%%%%%%%%%%%%%%%%%%%
%\begin{center} \small
%\begin{tabular}{|l|c|c|c|c|c|c|c|c|c|c|c|c|c|c|c|c|c|c|c|c|}
%\hline
%$T_L$ & \verb|fe| & \verb|ee| & \verb|ep| & \verb|ra| & \verb|rb| & \verb|rd| & \verb|ri| & \verb|ro| & \verb|al| & \verb|ar| & \verb|ba| & \verb|up| & \verb|pp| & \verb|de| & \verb|or| & \verb|he| & \verb|pe| & \verb|rr| & \verb|er| \\ 
%\hline
%$M_L$ &  &  &  &  &  &  &  &  &  &  &  &  &  &  &  &  &  &  & \\ 
%$(min.$ &  &  &  &  &  &  &  &  &  &  &  &  &  &  &  & \verb|he| & \verb|pe| & \verb|rr| & \verb|er| \\ 
%$motifs)$ &  &  &  &  &  &  &  &  &  &  &  &  &  &  &  &  &  &  & \\ 
% \hdashline[1pt/1pt]
%$M_L$ &  &  &  &  &  &  &  &  &  &  &  &  &  &  &  &  &  &  & \\ 
%$(rare$ & \verb|fe| & \verb|ee| &  &  & \verb|rb| & \verb|rd| &  &  &  &  &  & \verb|up| &  &  & \verb|or| &  &  &  & \verb|er| \\ 
%$motifs)$ &  &  &  &  &  &  &  &  &  &  &  &  &  &  &  &  &  &  & \\ 
%\hline
%\end{tabular}
%\end{center}
%\caption{Motif-Sets}~\label{tbl:Motif-Sets}
%%%%%%%%%%%%%%%%%%%%%%%%%%%%%%%%%%%%%%%%%%
%\end{table}

%%%%%%%%%%%%%%%%%%%%%%%%%%%%%%%%%%%%%
\end{example}

\subsection{Resolve-Sets}
The process of deducing the correct word from a motif occurrence can be compared to collision resolving of a hash-function. There may be many words mapped to the same motif. Upon encountering a motif, the match procedure must select the correct match (if there is a match) out of the specific word-set mapped to the motif. The following defines the hash-function and the set of words mapped to a single motif.

%%%%%%%%%%%%%%%%%%%%%%%%%%%%%%%%%%%%%
\begin{defn}[Motif-Set Hash Function]
	Any surjective function $H_{M_L}$ such that:
	\begin{eqnarray}
	  H_{M_L}&:&L\times \{0,1\} \BBsurjective M_L \ \ ,\ \nonumber  \\
	  H_{M_L}(w,i)=\mu & \Rightarrow & assoc_{i}(w,\mu)=1 \;  .
	\end{eqnarray}
\end{defn}
%%%%%%%%%%%%%%%%%%%%%%%%%%%%%%%%%%%%%

%%%%%%%%%%%%%%%%%%%%%%%%%%%%%%%%%%%%%
\begin{defn}[Resolve-Set]
	For a given motif $\mu\in M_L$ and hash-function $H_{M_L}$, $\mu$'s \itbf{Resolve-Set} $R_\mu\subseteq L$, is:
	\begin{eqnarray}
	  R_\mu:=\{w:H_{M_L}(w,0) = \mu \ \lor \ H_{M_L}(w,1) = \mu\} \;  .
	\end{eqnarray}
\end{defn}
%%%%%%%%%%%%%%%%%%%%%%%%%%%%%%%%%%%%%

\begin{example}[{{\rm Resolve-Sets}}]\label{ex:Resolve-Sets}
%%%%%%%%%%%%%%%%%%%%%%%%%%%%%%%%%%%%%
Resolve-sets for each one of the motifs in Example~\ref{ex:Motif-Sets} are shown in Table~\ref{tbl:Resolve-Sets}. Note that for a given motif-set, there may be many valid resolve-sets: for the \textit{minimum motifs} resolve-set in this example, instead of having an odd-motif mapping of \verb|ferrarri| to \verb|rr|, we can map this word to \verb|er| without changing the motif selection. See Section~\ref{subsec:Removing Duplicate Mappings} for details.

\begin{table}
\begin{center} \small
\begin{tabular}{|c|c|c|c|}
\hline
Resolve-Set & Resolve-Set & Resolve-Set & Resolve-Set \\ 
 Name & Contents & Contents & Contents \\ 
  & (minimum motifs) & (prefer rare motifs & (prefer rare motifs \\ 
  &  & in strings) & in input) \\ 
\hline
& & & \\[-8pt]
$R_{fe}$ &  & \verb|{ ferrarri }| &  \\
         &  & \verb|  ^^        | &  \\
 \hdashline[1pt/1pt]
$R_{ee}$ &  & \verb|{ deeper }| &  \\
         &  & \verb|   ^^     | &  \\
 \hdashline[1pt/1pt]
$R_{rb}$ &  & \verb|{ herbal }| & \verb|{ herbal }|  \\
         &  & \verb|    ^^    | & \verb|    ^^    |  \\
 \hdashline[1pt/1pt]
$R_{rd}$ &  & \verb|{ herd }| & \verb|{ herd }|  \\
         &  & \verb|    ^^  | & \verb|    ^^  |  \\
 \hdashline[1pt/1pt]
$R_{up}$ &  & \verb|{ upper }| &  \\
         &  & \verb|  ^^     | &  \\
 \hdashline[1pt/1pt]
$R_{or}$ &  & \verb|{ error }| &  \\
         &  & \verb|     ^^  | &  \\
 \hdashline[1pt/1pt]
$R_{he}$ & \verb|{ herd   ,| &  &  \\
         & \verb|  herbal }| &  &  \\
         & \verb|  ^^      | &  &  \\
 \hdashline[1pt/1pt]
$R_{pe}$ & \verb|{  upper ,| &  & \verb|{  upper ,|  \\
         & \verb|  deeper }| &  & \verb|  deeper }|  \\
         & \verb|     ^^   | &  & \verb|     ^^   |  \\
 \hdashline[1pt/1pt]
$R_{rr}$ & \verb|{    ferrarri ,| &  & \verb|{    ferrarri ,|  \\
         & \verb|      error   ,| &  & \verb|      error   ,|  \\
         & \verb|  ferrarri    }| &  & \verb|  ferrarri    }|  \\
         & \verb|       ^^      | &  & \verb|       ^^      |  \\
 \hdashline[1pt/1pt]
$R_{er}$ & \verb|{    herd   ,| & \verb|{    herd     ,| & \verb|{    herd   ,|  \\
         & \verb|     herbal ,| & \verb|     herbal   ,| & \verb|     herbal ,|  \\
         & \verb|   upper    ,| & \verb|   upper      ,| & \verb|   upper    ,|  \\
         & \verb|  deeper    ,| & \verb|  deeper      ,| & \verb|  deeper    ,|  \\
         & \verb|      error }| & \verb|      error   ,| & \verb|      error }|  \\
         & \verb|      ^^     | & \verb|     ferrarri }| & \verb|      ^^     |  \\
         &                      & \verb|      ^^       | &  \\
\hline
\end{tabular}
\end{center}
\caption{\itbf{Resolve-Sets:} The leftmost column lists all the motifs that belong to any of the 3 sets in Table~\ref{tbl:Motif-Sets}. The other 3 columns show, for each motif-selection scheme, the resolve-sets per motif. Two carets mark the motif position in each motif-set. Note that selecting the motif-set does not dictate the resolve-sets per motifs; for example, in the \textit{Minimum-Motifs} resolve-sets, \texttt{ferrarri} belongs to the \texttt{rr} resolve-set, but could also belong to the \texttt{er} resolve-set (these two options are interchangeable since both motifs appear at an odd offset within \texttt{ferrarri}).}~\label{tbl:Resolve-Sets}
\end{table}
%%%%%%%%%%%%%%%%%%%%%%%%%%%%%%%%%%%%%
\end{example}

%% file: b2-preprocessing.tex
\section{The Bouma2 Preprocessing Stage}\label{sec:The Bouma2 Preprocessing Stage}
As stated, Bouma2 essentially implements a hashing scheme for mapping the words in a language onto a motif-set. There are several approaches to constructing this mapping. The approach that is described in this paper relies on linear optimization, and consists of 3 steps:

\begin{enumerate}
\item Select an optimal motif-set out of the complete trace-set.
\item Remove duplicate mappings.
\item Build a \itbf{Mangled-Trie} for each motif.
\end{enumerate}

\subsection{Optimizing Motif Selection}\label{subsec:Optimizing Motif Selection}

The problem of selecting the best motif-set from a given trace-set can be formulated as a weighted \itbf{Integer Linear Programming}\cite{nemhauser1988integer} problem\footnote{In a previous paper~\cite{buchnik2011bouma2} we described how this problem is analogous to a \textit{Weighted Clique-Partitioning} problem.}. We use cost functions for motifs in order to optimize the selection for specific needs. Different cost functions serve different purposes, like improving performance, reducing memory footprint, or speeding up preprocessing time. Table~\ref{tbl:Motif-Sets} shows 3 different motif selections over the same trace-set.
%
%%%%%%%%%%%%%%%%%%%%%%%%%%%%%%%%%%%%%
\begin{defn}[The Bouma2 ILP Formulation]\label{def:The Bouma2 ILP Formulation}
Given a \itbf{Motif Cost Function} $c:\Sigma^2 \to \bbbr$, we define $M_{L,c}\subseteq T_{L}$ as a \itbf{Minimizing Motif Set} solving the following \textit{Integer Linear Programming} problem:
\begin{eqnarray}
  &\ &{\rm Minimize\ }\ \sum_{t\in T_L}c(t)x_t\ :\ \ \ x_t\in \{0,1\}\ \forall t\in T_L\ \nonumber \\
  &\ &{\rm subject\ to:\ }\ \forall w\in L:\ \nonumber \\
  &\ &\sum_{t\in T_{L}}x_t\cdot assoc_{0}(w,t) \geq 1\ \land\ \sum_{t\in T_{L}}x_t\cdot assoc_{1}(w,t)\geq 1 \;  .
\end{eqnarray}
\end{defn}
%%%%%%%%%%%%%%%%%%%%%%%%%%%%%%%%%%%%%
%
\begin{example} [{{\rm Minimizing Motif False-Positives}}]\label{ex:Minimizing Motif False-Positives}
%%%%%%%%%%%%%%%%%%%%%%%%%%%%%%%%%%%%%
Cost functions can facilitate the use of statistics gathered on the input string and language. For example, the conditional probability $P(w \mid t)$, \ie the probability of the word $w$ appearing in the input string, given that the trace $t$ was observed, can be used as a weight (note that here we need to \itbf{maximize} the conditional probability, so we negate the cost function):
\begin{equation}
  c(t)=-\sum_{w\in L}(assoc_{0}(w,t)\lor assoc_{1}(w,t)) \cdot P(w \mid t) \;  .
\end{equation}
%%%%%%%%%%%%%%%%%%%%%%%%%%%%%%%%%%%%%
\end{example}

\begin{example} [{{\rm Memory Cost Function}}]
%%%%%%%%%%%%%%%%%%%%%%%%%%%%%%%%%%%%%
Subject to proper implementation, minimizing the number of motifs may help reduce overall memory requirements:
\begin{equation}
  c(t)=1 \;  .
\end{equation}
%%%%%%%%%%%%%%%%%%%%%%%%%%%%%%%%%%%%%
\end{example}

\subsection{Removing Duplicate Mappings}\label{subsec:Removing Duplicate Mappings}

The motif-selection process may not provide a unique mapping of words to motifs: for a given motif-set, the same word may be mapped to more than 2 motifs in the set. We need to make sure that each word has only two mappings\footnote{This rule has exceptions: \eg for case-insensitive text matches (see~\ref{subsec:Pandemonium}), Bouma2 can accept up to 8 mappings per word.}, otherwise the match-time algorithm would generate duplicate reports for the same match. Redundant mappings may be to distinct motifs or to the same motif. For example, the string \verb|http://www.wwwdotcom.com| may be mapped 4 times to \verb|ww|: twice to an even offset and twice to an odd offset. Also observe Table~\ref{tbl:Resolve-Sets}: for the motif selection with minimum motifs, the word \verb|ferrarri| could be either mapped to \verb|rr| or to \verb|er| - both are legitimate odd motif mappings (although note the difference in the depth of the resulting mangled-tries - see Section~\ref{sec:The Match Process}). When selecting which mapping to remove, we consider symbol occurrence statistics, complexity of the resulting mangled-tries, relative offset of motifs etc.

\subsection{Mangled-Trie Construction}\label{subsec:Mangled-Trie Construction}

When several pattern strings map to the same motif, upon finding this motif the match procedure has to decide how many of them (zero or more) actually match, and also report the exact match position for every matching string. For this purpose we have developed \itbf{the Mangled-Trie} data-structure. The mangled-trie is a special decision-tree, which \itbf{dictates} the next position to examine in the input, while proceeding along any one of its branches. This is opposed to a regular \textit{trie}\citeb{fredkin1960trie}, which proceeds in a predefined consume-order (evidently through the use of a \textit{Consume-Order Dependent} algorithm). Essentially, the mangled-trie is built over an extended set of symbols, that includes the relative offset dimension. This allows us to predefine the sequence of examined symbols as offsets relative to the motif match.

%%%%%%%%%%%%%%%%%%%%%%%%%%%%%%%%%%%%%%%%%%%%
\begin{defn}[The Mangled-Trie Symbol Set]\label{defn:The Mangled-Trie Symbol Set}
	For a given language $L$, motif $\mu \in M_L$ and resolve-set $R_\mu \subseteq L$, the corresponding \itbf{Mangled-Trie Symbol Set} is:
	\begin{eqnarray}
	  &{\Sigma_{{M.T.}^{L,\mu}}}& = \{(w, i): w \in R_\mu, -l^{w,\mu} \leq i < |w| - l^{w,\mu}, i \neq 0, i \neq 1\}  \nonumber  \\
	  &{\rm where }& l^{w,\mu} {\rm \ \ satisfies: \ \ } occ(w, \mu, l^{w,\mu}) = 1  \; .
	\end{eqnarray}	
\end{defn}
%%%%%%%%%%%%%%%%%%%%%%%%%%%%%%%%%%%%%%%%%%%%

The different types of mangled-trie nodes can be distinguished according to the following terminology:
\begin{enumerate}
\item \itbf{State:} This node is associated with a single offset relative to the motif position. It maps transitions to child nodes along the mangled-trie according to the symbol value at that offset. Any path along the mangled-trie may include no more than one state per offset.
\item \itbf{Transitional:} This node is a special type of state: it indicates a match of a single string that resulted from consuming the symbol during the last transition. No further matches are required for this specific string, but if it is a substring of any other string in the mangled-trie that was not ruled out yet, the match proceeds.
\item \itbf{Terminal:} This node holds information about remaining fragments of a \itbf{single} string, whereas all the other strings were already ruled out while traversing the mangled-trie. If all the fragments match, this node produces a successful match result for its associated string.
\item \itbf{Pivot:} This node indicates a \quotes{fallback} state, that has to be visited after the current path is consumed. This occurs when there are two disjoint sets of string fragments - fragments that reside to the left of the motif position, and fragments that reside to the right, with no single string having fragments on both sides. Any of the above node types may also be a pivot. Also, any of the above node types may proceed to a pivot, but there cannot be more than one pivot along a single mangled-trie path.
\end{enumerate}

The heuristic described in~\ref{sec:Mangled-Trie Construction Algorithm} constructs a mangled-trie from a given resolve-set. It recursively selects a \quotes{scoring} offset, and examines the possible \singlequote{interesting} symbol values at that offset. Every unique symbol in the scoring offset is considered separately, given that if this symbol would be found in the input at that offset, it will allow us to discard all the words that do not match this symbol. A subtrie is constructed for the remaining symbols belonging to the remaining words.

%%%%%%%%%%%%%%%%%%%%%%%%%%%%%%%%%%%%%%%%%%%%
\begin{thm}[Maximum Mangled-Trie Depth]\label{thm:Maximum Mangled-Trie Depth}
	Let $w^{max}$ be the longest word in $L$ ($|w^{max}| \geq w, \forall w \in L$). Then the depth of any mangled-trie over $L$ satisfies:
	\begin{equation}
	  depth({M.T.}^{L,\mu}) \leq 2 \cdot (|w^{max}| - 2) \;  .
	\end{equation}
	\begin{pf}
		Every offset that is examined against a mangled-trie state allows us to eliminate all the strings that do not match its actual value. Therefore, if we have more offsets with overlapping strings, we can eliminate strings in fewer steps and our mangled-trie will be shallower. The worst case is when we have no overlapping offsets at all (except for the motif position itself, at offsets $0$ and $1$): 
		\begin{eqnarray}
		  \forall w \in R_\mu &:& occ(w, \mu, |w|-2)=1  \land (w,2) \notin {\Sigma_{{M.T.}^{L,\mu}}} \nonumber  \\
		  \ &\lor &\ occ(w, \mu, 0)=1 \land (w,-1) \notin {\Sigma_{{M.T.}^{L,\mu}}}\;  .
		\end{eqnarray}
		Obviously, in this case we would have to examine the input both before the motif match and after it, no more than $|w^{max}| - 2$ symbols in each direction.
		\qed
	\end{pf}
\end{thm}
%%%%%%%%%%%%%%%%%%%%%%%%%%%%%%%%%%%%%%%%%%%%

\begin{example}[{{\rm Mangled-Trie Construction}}]\label{ex:Mangled-Trie Construction}
%%%%%%%%%%%%%%%%%%%%%%%%%%%%%%%%%%%%%
The following example describes the recursive construction of a mangled-trie out of a given resolve-set. Assume that the \quotes{scoring} offset is given to us at every iteration\footnote{Choosing different scoring offsets would generate different mangled-tries, possibly giving room for further optimization of memory consumption or mangled-trie depth.}. Consider $R_{er}$ for the minimum-motifs option in Table~\ref{tbl:Resolve-Sets} (note that we strike through the symbols at the motif position, to emphasize that they will not be considered when building the mangled-trie):
\needspace{6\baselineskip}
\begin{quote}
\verb|{    h|\st{\texttt{er}}\verb|d   ,| \\
\verb|     h|\st{\texttt{er}}\verb|bal ,| \\
\verb|   upp|\st{\texttt{er}}\verb|    ,| \\
\verb|  deep|\st{\texttt{er}}\verb|    ,| \\
\verb|      |\st{\texttt{er}}\verb|ror }| \\
\verb|      ^^     |
\end{quote}
Assume that the first scoring offset is -1. At this offset, there are only 3 \singlequote{interesting} possibilities of symbol values: \verb|h|, \verb|p| or any other symbol. We examine each possibility separately: if \verb|h| is found at offset -1, then all strings requiring \verb|p| at this offset (\ie \verb|upper| and \verb|deeper|) can be \singlequote{purged}. If \verb|p| is found, then all strings requiring \verb|h| (\ie \verb|herd| and \verb|herbal|) are purged. If any other symbol is found, we can purge all strings that require a specific symbol at offset -1, \ie strings containing either \verb|h| or \verb|p| (specifically, \verb|herd|, \verb|herbal|, \verb|upper| and \verb|deeper|). Adhering to the notation in Algorithm~\ref{alg:B2-BUILD-SUBTRIE}, we specify the set of unique symbols for offset -1 as: $A \leftarrow \{\verb|h,p|\}$. We first purge by \verb|h| (the '\verb|.|' sign specifies a symbol that is ignored either because it belongs to a purged string, or because the offset it resides in was already examined):
\needspace{6\baselineskip}
\begin{quote}
\verb|{    .|\st{\texttt{er}}\verb|d   ,| \\
\verb|     .|\st{\texttt{er}}\verb|bal ,| \\
\verb|   .....    ,| \\
\verb|  ......    ,| \\
\verb|      |\st{\texttt{er}}\verb|ror }| \\
\verb|      ^^     |
\end{quote}
We are now left with the strings \verb|erd|, \verb|erbal| and \verb|error|. We will now recursively create a \singlequote{sub-mangled-trie} that would specifically resolve matches for these strings. For the purged subtrie, assume the scoring offset is 2, giving $A \leftarrow \{\verb|d,b,r|\}$. Encountering any one of these symbols at offset 2 would cause the elimination of all the words that do not contain it (\eg \verb|b| eliminates \verb|erd| and \verb|error|). Encountering \verb|d| successfully terminates the match without requiring further validation, reporting \verb|herd|. We therefore add a \itbf{Transitional} node for \verb|d|. \verb|b| still requires extra substring matching, so we add a \itbf{Terminal} for \verb|b|, validating \verb|al| at offset 3. Similarly, we add a \itbf{Terminal} for \verb|r|, validating \verb|or| at offset 3. Note that if the input at offset 2 contains any symbol other than \verb|d,b| or \verb|r|, we conclude with no match.
Returning to the root state, we now purge by \verb|p|:
\needspace{6\baselineskip}
\begin{quote}
\verb|{    ....   ,| \\
\verb|     ...... ,| \\
\verb|   up.|\st{\texttt{er}}\verb|    ,| \\
\verb|  dee.|\st{\texttt{er}}\verb|    ,| \\
\verb|      |\st{\texttt{er}}\verb|ror }| \\
\verb|      ^^     |
\end{quote}
We are left with the following string fragments: \verb|up| at offset -3 for \verb|upper|, \verb|dee| at offset -4 for \verb|deeper|, and \verb|ror| at offset 2 for \verb|error|. We thus have two disjoint subsets of words, each one on a different side of the motif match. If the next scoring offset is chosen to the left (\ie either -4, -3 or -2), then the resulting subtrie would handle only the fragments on the left, and its \singlequote{leaves} would be followed by a \itbf{Pivot State} for handling the resolving of the right side. Alternatively, if the scoring offset is chosen on the right (\ie either 2, 3 or 4) then a subtrie would handle the single remaining fragment on the right, and would be followed by a \itbf{Pivot State} that would handle the left side. Assume the scoring offset this time is -2. $A \leftarrow \{\verb|p,e|\}$ for this offset, and again each symbol choice uniquely identifies one of the two words on the left.This allows us to add the corresponding \itbf{Terminals} (\verb|u| at offset -3 if \verb|p| is found, and \verb|de| at offset -4 if \verb|e| is found) in order to complete the match. But this time, no matter if we find a match on the left side or not, we also have to match another \itbf{Terminal} on the right side (namely \verb|ror| at offset 2). This \itbf{Terminal} is added as a \itbf{Pivot State}.
Finally, we return once more to the root state at offset -1 and purge for the \quotes{fallback} case (\ie neither \verb|h| nor \verb|p|):
\needspace{6\baselineskip}
\begin{quote}
\verb|{    ....   ,| \\
\verb|     ...... ,| \\
\verb|   .....    ,| \\
\verb|  ......    ,| \\
\verb|      |\st{\texttt{er}}\verb|ror }| \\
\verb|      ^^     |
\end{quote}
We are left with a single fragment: \verb|ror| at offset 2 for matching \verb|error|. We need to add a \itbf{Terminal} for matching this fragment, but it is identical to the \itbf{Pivot State} we added earlier, so the two nodes can be consolidated. The complete mangled-trie is shown in Figure~\ref{fig:Mangled-Trie}.
\begin{figure}
	\centering
	\includegraphics[width=0.95\textwidth]{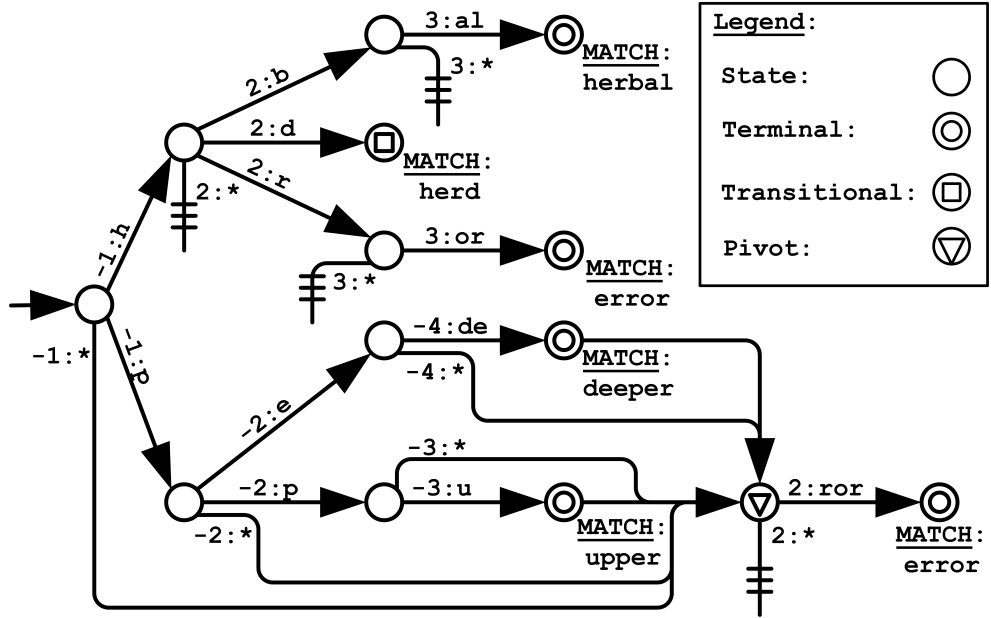}
	\caption{Mangled-Trie for $R_{er}$ in Example~\ref{ex:Mangled-Trie Construction}.}
	\label{fig:Mangled-Trie}
\end{figure}
%
%%%%%%%%%%%%%%%%%%%%%%%%%%%%%%%%%%%%%
\end{example}

%% file: b2-match-process.tex
\section{The Match Process}\label{sec:The Match Process}
The preprocessing stage constructs two separate data-structures: a map for searching motif occurrences and a set of mangled-tries for resolving motif matches. It is beneficial to separate motif finding from motif resolving, because in practical applications this allows reuse of cached data. We adopt the notion of \itbf{Fast-Path} and \itbf{Slow-Path} common in networking (see \cite{govindan2002estimating}) to differentiate between the time-critical, deterministic process of motif finding and the input-sensitive process of motif resolving, respectively. ~\ref{sec:Match-Process Algorithm} presents the match-time procedures.

%%%%%%%%%%%%%%%%%%%%%%%%%%%%%%%%%%%%%%%%%%%%
\begin{thm}[Bouma2 Memory Consumption]\label{thm:Bouma2 Memory Consumption}
	For a given language $L$ and a motif-set $M_L$, the amount of memory required for the Bouma2 match structures is $O(4 \cdot |L| \cdot (|w^{max}| - 2) + |M_L|)$.
	\begin{pf}
		The Fast-Path procedure queries data based on the motif-set, hence $|M_L|$. For the Slow-Path, every word is represented by two separate paths along one or two mangled-tries. By Theorem~\ref{thm:Maximum Mangled-Trie Depth}, such a path would consume up to $O(2 \cdot (|w^{max}| - 2))$ memory, hence the result above.
		\qed
	\end{pf}
\end{thm}
%%%%%%%%%%%%%%%%%%%%%%%%%%%%%%%%%%%%%%%%%%%%

%%%%%%%%%%%%%%%%%%%%%%%%%%%%%%%%%%%%%%%%%%%%
\begin{thm}[Bouma2 Worst-Case Complexity]\label{thm:Bouma2 Worst-Case Complexity}
	Let the aggregate occurrence probability of any of the motifs in $M_L$ be $P_{M_L}=\sum_{\mu \in M_L}P_\mu$. Then the worst-case complexity of the Bouma2 match procedure over an input of length $n$ is:
	\begin{equation}
	  O(n \cdot (0.5 + P_{M_L} \cdot (|w^{max}|-2))) \;  .
	\end{equation}
	\begin{pf}
		During Fast-Path, $n/2$ comparisons are made over the input. Then the number of motif matches is $P_{M_L} \cdot n/2$. In the worst case, the Slow-Path would traverse the full depth of the mangled-tries for every motif occurrence, so the number of Slow-Path operations according to Theorem~\ref{thm:Maximum Mangled-Trie Depth} is:
		\begin{equation}
		  (P_{M_L} \cdot n/2) \cdot (2 \cdot (|w^{max}| - 2)) \;  .
		\end{equation}
		The sum of the Fast-Path operations and the Slow-Path operations yields the result above.
		\qed
	\end{pf}
\end{thm}
%%%%%%%%%%%%%%%%%%%%%%%%%%%%%%%%%%%%%%%%%%%%

Theorem~\ref{thm:Bouma2 Worst-Case Complexity} demonstrates the importance of optimizing the motif-set quality: for a given word-set, we can improve the worst-case match performance simply by selecting a better set of motifs.

%% file: b2-experimental-results.tex
\section{Experimental Results}\label{sec:Experimental Results}
\itbf{Snort}\reg \footnote{\url{http://www.snort.org}} is a popular Open-Source \textit{Intrusion Prevention System} maintained by SourceFire, Inc. It is an excellent case-study for extensive use of multiple exact string-match: Snort uses Aho-Corasick for initial fast filtering of cyber-attack signatures. Although the Aho-Corasick variant that is built into Snort was originally introduced as a performance optimization, it is in itself one of the largest bottlenecks of this software (see \citeb{norton2004optimizing}).

It is claimed in \citeb{tuck2004deterministic} that up to $70\%$ of the Snort execution time is spent on various string-matching algorithms. In the experiments we conducted against Internet Service Provider packet captures, the default Snort Aho-Corasick variant accounted for most of the string-matching overhead, taking around $40\%$ of the total execution time.

Snort uses Aho-Corasick as follows: it builds several unique AC state-machines, based on the various TCP/UDP port groups specified in any of its rules, and an extra one for rules that are not port-specific. This allows initial port-based traffic filtering, and also probably reduces the total memory consumption and helps in localizing memory accesses compared with a single AC for all strings. On the other hand, this means that some strings may be accounted for in several distinct AC instances. Nevertheless, we test the AC against a single Bouma2 instance (which is inherently localized), with a single representation of each unique string (together with a reference to its duplicates). It should also be mentioned that some of the AC instances (around 3.5\% of the entire set) are built to perform 2-symbol strides. We specify the number of AC instances built in each test.

We used Snort v2.9.1.1 source code for Windows. For rules we used the v2.9.1.1 rule-set released on Oct. 6\textsuperscript{th} 2011. In order to comply with the Bouma2 version under test, which accepts only strings 3 bytes long or more (see~\ref{subsec:Short Strings} for a discussion on short-string support), we identified and disabled all Snort rules that contain 2-byte and 1-byte strings. In the Snort code we implanted calls to the Bouma2 API, to allow running the 3 Bouma2 variants on the same input that AC was receiving. The Bouma2 preprocessor and matcher were written in C++ using \textit{Microsoft\reg Visual Studio\reg 2010 Premium}. The motif selection process was implemented with source-code from the COIN-OR\citeb{lougee2003common} BCP\citeb{Mar03} project\footnote{\url{https://projects.coin-or.org/Bcp}}.

All tests were done on a \textit{Dell\texttrademark} computer with \textit{Intel\reg Core\texttrademark 2 Duo CPU 2.53 GHz} with \textit{1.95 GB RAM}, running \textit{Windows XP SP3}. In our tests we use the \textit{Microsoft\reg Visual Studio\reg 2010 Premium Sampling Profiler}. We conducted 5 different tests, using the default Snort rule-set and increasing the number of enabled rules with every test. As input we used a packet capture of traffic sampled at a large Internet Service Provider site. We applied the packets to Snort using the \verb|-r| option. For the \textit{Rare Motifs in Input} variant we used statistics gathered on one third of the entire capture. Match results were verified to be identical in a dry run.

We compare performance by means of \itbf{Algorithm Throughput}, taking into account the processor frequency (2.53 GHz), the sample interval (every 10,000,000 clock cycles), the number of bytes sent to the match procedure (around 1 GByte, actual value calculated during the test) and finally the number of samples recorded by the profiler for each match procedure. We use Equation~\ref{eqn:Algorithm Throughput}:

\begin{eqnarray}\label{eqn:Algorithm Throughput}
	SampleInterval & = & \frac{10,000,000}{2.53 \cdot 1,000} = 3952.569 \mu sec. \nonumber  \\
	Throughput & = & \frac{BytesConsumed \cdot 8}{SampleCount \cdot SampleInterval}  \nonumber  \\
	& = & 0.002024 \cdot \frac{BytesConsumed}{SampleCount} Mbits/sec  \;  .
\end{eqnarray}

We built 3 different versions of Bouma2, with 3 different cost functions (see Definition~\ref{def:The Bouma2 ILP Formulation}):
\begin{enumerate}
\item \itbf{Minimum Motifs}: $c(t)=1$
\item \itbf{Rare Motifs in Strings\footnote{In the 5th test, the BCP algorithm failed to find a solution for the \textit{Rare-Motifs-in-Strings} variant in reasonable time.}}: $c(t)=\sum_{w\in L}\sum_{0 \leq l < |w|}occ(w,t,l)$
\item \itbf{Rare Motifs in Input} (by occurrence probabilities): $c(t)=P(t)$
\end{enumerate}

The results are presented in Tables~\ref{tbl:Snort Benchmark (657 strings)},~\ref{tbl:Snort Benchmark (1,751 strings)},~\ref{tbl:Snort Benchmark (2,443 strings)},~\ref{tbl:Snort Benchmark (4,949 strings)} and~\ref{tbl:Snort Benchmark (7,146 strings)}. Note the similarity between the occurrence probabilities calculated from the statistics that were gathered on one third of the traffic, and the actual numbers collected during the match.

\begin{table}[ht]
%%%%%%%%%%%%%%%%%%%%%%%%%%%%%%%%%%%%%%%%%%
\begin{center} \small
\begin{tabular}{|c|c|c|c|c|c|c|}
\hline
 Test   & Word    & Unique  & Traces & Words Agg.   & Unique Words      & Snort AC  \\ 
 No.    & Count   & Words   &        & Size (bytes) & Agg. Size (bytes) & Instances \\ 
\hline
  1     &     657 &    578  & 1,742  &       6,966  &             6,301 & 201  \\ 
\hline
\end{tabular}
\begin{tabular}{|l|r|r|c|c|c|}
\hline
       & Throughput   & Memory   & Motifs & Motif Occur. & Motif Occur.  \\ 
       & (MBits/sec)  & (bytes)  &        & Prob. (est.)     & Prob. (actual)  \\ 
\hline
AC     &  1,877.544543 & 2,550,000  & &                  &   \\ 
B2-M   &  2,972.778859 &  524,800  & 254 & 0.0241563        & 0.028288445  \\ 
B2-RS  &  3,086.110605 &  539,392  & 354 & 0.0221357        & 0.024200245  \\ 
B2-RI  &  3,513.735279 &  525,312  & 309 & 0.0190495        & 0.020905283  \\ 
\hline
\end{tabular}
\end{center}
\caption{Snort Benchmark (657 strings)}~\label{tbl:Snort Benchmark (657 strings)}
%%%%%%%%%%%%%%%%%%%%%%%%%%%%%%%%%%%%%%%%%%
\end{table}

\begin{table}
%%%%%%%%%%%%%%%%%%%%%%%%%%%%%%%%%%%%%%%%%%
\begin{center} \small
\begin{tabular}{|c|c|c|c|c|c|c|}
\hline
 Test   & Word    & Unique  & Traces & Words Agg.   & Unique Words      & Snort AC  \\ 
 No.    & Count   & Words   &        & Size (bytes) & Agg. Size (bytes) & Instances \\ 
\hline
  2     &   1,751 &  1,290  &  3,655 &      24,062  &            17,349 & 285  \\ 
\hline
\end{tabular}
\begin{tabular}{|l|r|r|c|c|c|}
\hline
       & Throughput   & Memory   & Motifs & Motif Occur. & Motif Occur.  \\ 
       & (MBits/sec)  & (bytes)  &        & Prob. (est.)     & Prob. (actual)  \\ 
\hline
AC     &  1,594.655819 & 9,090,000  & &                  &   \\ 
B2-M   &  2,321.656187 &  1,168,896  & 396 & 0.0383088        & 0.037200262  \\ 
B2-RS  &  2,064.91349  &  1,138,432  & 684 & 0.0361115        & 0.035557825  \\ 
B2-RI  &  2,543.281109 &  1,181,952  & 487 & 0.0304228        & 0.030030249  \\ 
\hline
\end{tabular}
\end{center}
\caption{Snort Benchmark (1,751 strings)}~\label{tbl:Snort Benchmark (1,751 strings)}
%%%%%%%%%%%%%%%%%%%%%%%%%%%%%%%%%%%%%%%%%%
\end{table}

\begin{table}
%%%%%%%%%%%%%%%%%%%%%%%%%%%%%%%%%%%%%%%%%%
\begin{center} \small
\begin{tabular}{|c|c|c|c|c|c|c|}
\hline
 Test   & Word    & Unique  & Traces & Words Agg.   & Unique Words      & Snort AC  \\ 
 No.    & Count   & Words   &        & Size (bytes) & Agg. Size (bytes) & Instances \\ 
\hline
  3     &   2,443 &  1,609  &  4,041 &      31,843  &            22,275 & 336  \\ 
\hline
\end{tabular}
\begin{tabular}{|l|r|r|c|c|c|}
\hline
       & Throughput   & Memory   & Motifs & Motif Occur. & Motif Occur.  \\ 
       & (MBits/sec)  & (bytes)  &        & Prob. (est.)     & Prob. (actual)  \\ 
\hline
AC     &  1,433.464052  & 11,480,000  & &                  &   \\ 
B2-M   &  1,855.21453   &  1,531,392  & 460 & 0.0405084        & 0.039606471  \\ 
B2-RS  &  2,186.104496  &  1,497,088  & 761 & 0.0390466        & 0.038530029  \\ 
B2-RI  &  2,239.971548  &  1,507,328  & 547 & 0.0327172        & 0.032419297  \\ 
\hline
\end{tabular}
\end{center}
\caption{Snort Benchmark (2,443 strings)}~\label{tbl:Snort Benchmark (2,443 strings)}
%%%%%%%%%%%%%%%%%%%%%%%%%%%%%%%%%%%%%%%%%%
\end{table}

\begin{table}
%%%%%%%%%%%%%%%%%%%%%%%%%%%%%%%%%%%%%%%%%%
\begin{center} \small
\begin{tabular}{|c|c|c|c|c|c|c|}
\hline
 Test   & Word    & Unique  & Traces & Words Agg.   & Unique Words      & Snort AC  \\ 
 No.    & Count   & Words   &        & Size (bytes) & Agg. Size (bytes) & Instances \\ 
\hline
  4     &   4,949 &  3,296  &  7,795 &      77,282  &            54,475 & 414  \\ 
\hline
\end{tabular}
\begin{tabular}{|l|r|r|c|c|c|}
\hline
       & Throughput   & Memory   & Motifs & Motif Occur. & Motif Occur.  \\ 
       & (MBits/sec)  & (bytes)  &        & Prob. (est.)     & Prob. (actual)  \\ 
\hline
AC     &  1,052.746683  & 28,600,000  & &                  &   \\ 
B2-M   &  1,679.054087  &  3,259,392  & 705  & 0.0469979        & 0.045910587  \\ 
B2-RS  &  1,697.318856  &  3,280,640  & 1,058 & 0.047068         & 0.046318193  \\ 
B2-RI  &  1,763.07455   &  3,266,048  & 814  & 0.0410833        & 0.040417134  \\ 
\hline
\end{tabular}
\end{center}
\caption{Snort Benchmark (4,949 strings)}~\label{tbl:Snort Benchmark (4,949 strings)}
%%%%%%%%%%%%%%%%%%%%%%%%%%%%%%%%%%%%%%%%%%
\end{table}

\begin{table}
%%%%%%%%%%%%%%%%%%%%%%%%%%%%%%%%%%%%%%%%%%
\begin{center} \small
\begin{tabular}{|c|c|c|c|c|c|c|}
\hline
 Test   & Word    & Unique  & Traces & Words Agg.   & Unique Words      & Snort AC  \\ 
 No.    & Count   & Words   &        & Size (bytes) & Agg. Size (bytes) & Instances \\ 
\hline
  5     &   7,146 &  4,841  & 8,789  &     131,547  &            98,546 & 424  \\ 
\hline
\end{tabular}
\begin{tabular}{|l|r|r|c|c|c|}
\hline
       & Throughput   & Memory   & Motifs & Motif Occur. & Motif Occur.  \\ 
       & (MBits/sec)  & (bytes)  &        & Prob. (est.)     & Prob. (actual)  \\ 
\hline
AC     &  841.9984458  & 51,370,000    & &                  &   \\ 
B2-M   &  1,498.041131  &  4,859,136    & 862  & 0.0520728   & 0.050279535  \\ 
B2-RS  &  -            &  -          & -    & -           & -  \\ 
B2-RI  &  1,697.08156   &  4,861,184    & 985  & 0.0455084   & 0.044300929  \\ 
\hline
\end{tabular}
\end{center}
\caption{Snort Benchmark (7,146 strings)}~\label{tbl:Snort Benchmark (7,146 strings)}
%%%%%%%%%%%%%%%%%%%%%%%%%%%%%%%%%%%%%%%%%%
\end{table}

\begin{figure}[ht]
	\centering
	\includegraphics[width=0.98\textwidth]{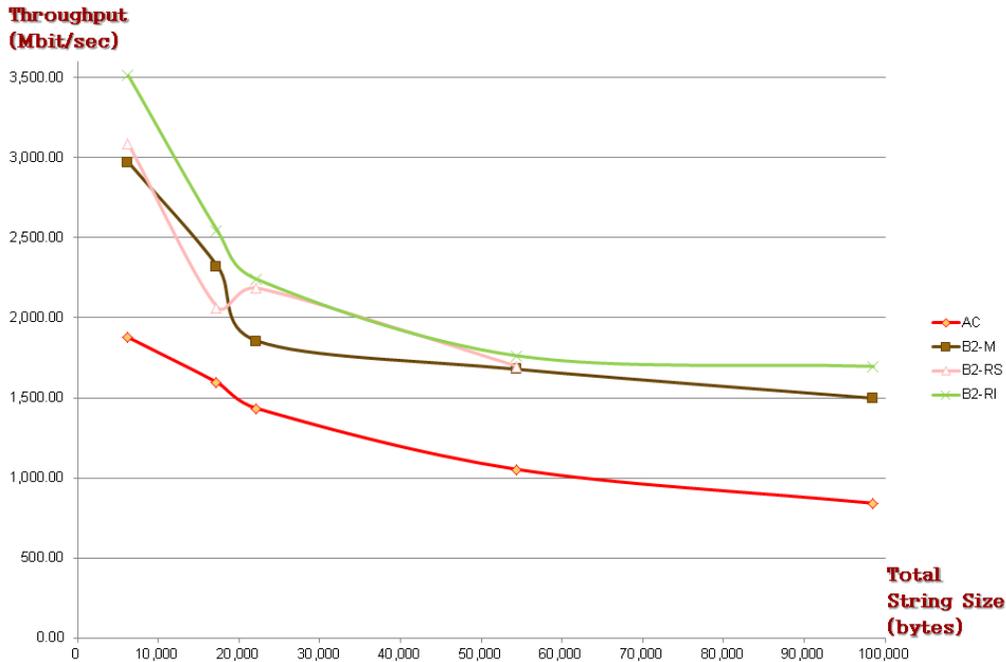}
	\caption{Throughput comparison for tests in Section~\ref{sec:Experimental Results}.}
	\label{fig:Throughput Comparison}
\end{figure}

\begin{figure}[ht]
	\centering
	\includegraphics[width=0.98\textwidth]{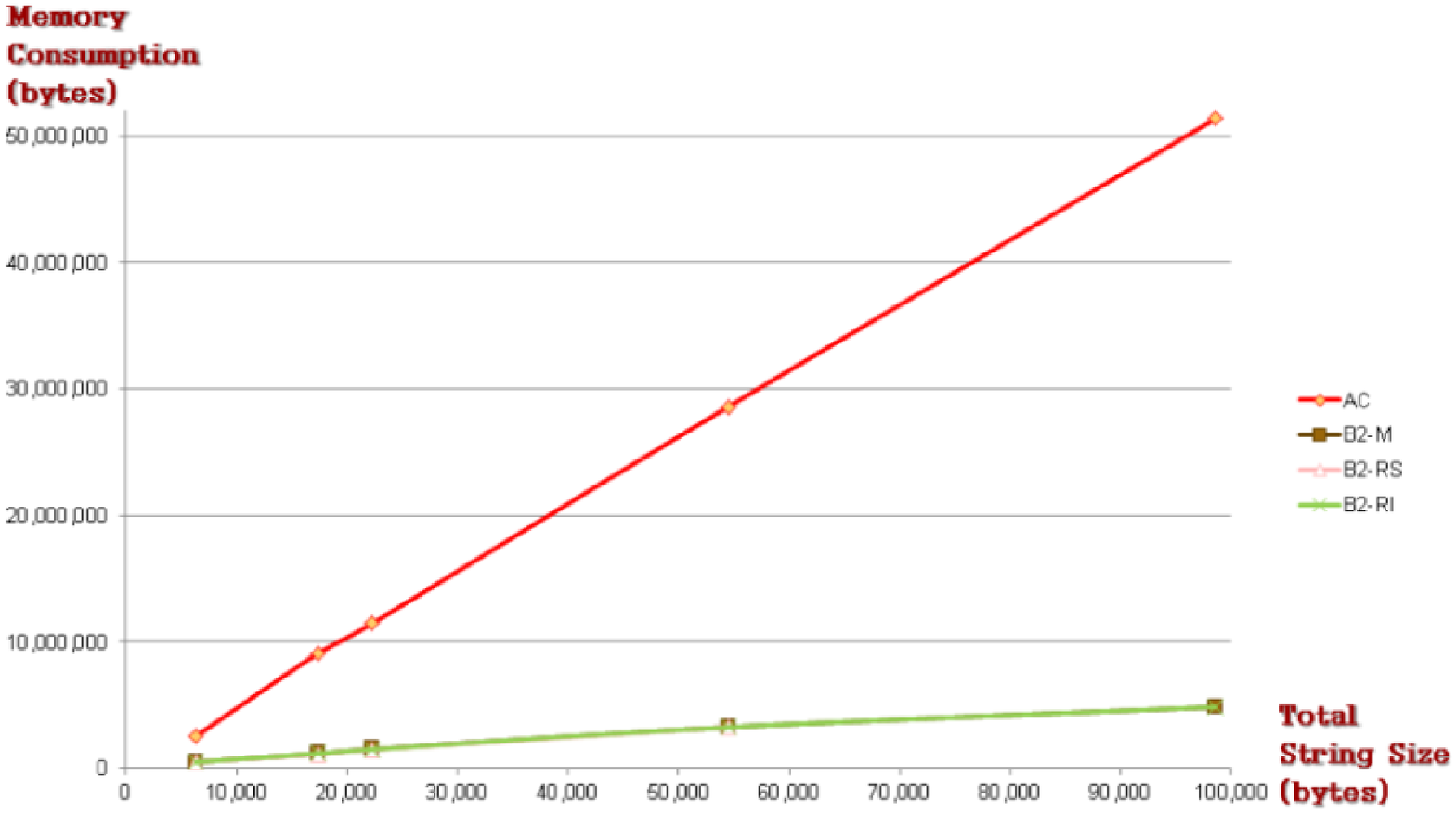}
	\caption{Memory comparison for tests in Section~\ref{sec:Experimental Results}.}
	\label{fig:Memory Comparison}
\end{figure}

The throughput comparison is visualized in Figure~\ref{fig:Throughput Comparison}. It is evident that Bouma2 achieves around double the throughput compared with AC. Moreover, choosing the Bouma2 version that is suitable for the input may improve the throughput: preferring motifs that are rare in the input achieves a 13\% improvement compared with the minimum-motifs version. A comparison of the memory consumption is displayed in Figure~\ref{fig:Memory Comparison}. Evidently, Bouma2 requires 10 times less memory than the Snort version of AC.

%%%%%%%%%%%%%%%%%%%%%%%%%%%%%%%%%%%%%%%%%%
%%%%%%%%%%%%%%%%%%%%%%%%%%%%%%%%%%%%%%%%%%

%%%%%%%%%%%%%%%%%%%%%%%%%%%%%%%%%%%%%%%%%%
%%%%%%%%%%%%%%%%%%%%%%%%%%%%%%%%%%%%%%%%%%

%% file: b2-conclusion.tex
\section{Conclusion}\label{sec:Conclusion}
In this paper we presented Bouma2, which is to our knowledge the first \itbf{Consume-Order Agnostic} multiple exact string-match algorithm. This approach allows independent (and therefore parallelizable) examination of different segments of the input, and as such corresponds well with modern processor and other hardware architectures, which allow fast random access to a fairly large amount of cached memory. This is opposed to the \textit{Consume-Order Dependent} State-Machine model, which assumes access to a single symbol at a time, and as such corresponds better with the theoretical single-tape \textit{Turing-Machine} \citeb{Turing-1936} concept.

One lesson that we learn from the tests in Section~\ref{sec:Experimental Results} may seem obvious: \itbf{real-life data is not random}. We can (and should) rely on premature knowledge of data characteristics when applying operations on new data. Different types of data have different characteristics, and an adaptive version of Bouma2 will be able to learn these characteristics and improve its choice of motifs on-the-fly.

We believe that regarding the problem of improving pattern-match performance as a linear optimization problem sheds new light on this well-researched area. The Branch-and-Cut algorithm has proven itself in our case as a powerful and flexible tool that should be explored and used more. Taking Bouma2 as a case-study and applying the same approach to other areas of study may also prove beneficial. As an example we give the work in \citeb{DBLP:conf/infocom/Bremler-BarrDHK12}, which is trying to address very recent needs of \textit{Deep-Packet-Inspection} over compressed input by giving a solution that is tightly coupled with the Aho-Corasick scheme. An alternative \textit{Consume-Order Agnostic} solution applying linear optimization for both compression and DPI may also be considered.

%% file: b2-future-work.tex
\section{Future Work}\label{sec:Future Work}
\subsection{Pandemonium}\label{subsec:Pandemonium}

\itbf{Pandemonium}\footnote{The term was chosen as a tribute to Selfridge's cognitive \textit{Pandemonium Model} \citeb{Selfridge58}.} is a regular expression matching library, which is powered by Bouma2 and follows similar concepts. As with exact string-match, regular expression match can also be \textit{Consume-Order Agnostic}: in essence, a regular expression is a set of queries over the input text; \itbf{ALL} the queries have to yield a positive result for the match to succeed, regardless of the order in which they are performed. In many cases, matches can be abandoned at an early stage based on negative query results, which may have been obtained by efficiently leveraging match results from Bouma2. Pandemonium can utilize Bouma2's Fast-Path match results to perform advanced matches (\eg case-insensitive match of specific words). Another powerful feature is that Pandemonium can accept multiple regular expressions and match them in parallel.

\subsection{Performance Improvements}\label{subsec:Performance Improvements}

It is difficult to estimate the nominal impact of improving the motif-set on performance. Since the actual Slow-Path performance depends on the specific mangled-tries that are accessed, the weight applied when selecting a motif should also be affected by the relative complexity of the resulting mangled-trie. We are thus researching methods of improving the motif cost functions in this manner (\eg just like trace occurrence counts, motif-specific performance data for evaluating mangled-trie costs can also be collected on-the-fly).

\subsection{Complexity}\label{subsec:Complexity}

It would be beneficial to refine the complexity result obtained in Theorem~\ref{thm:Bouma2 Worst-Case Complexity}, and specifically find bounds that do not rely on a given motif-set selection, but rather on the diversity of trace values in the word-set. Also, finding bounds for the best-case complexity and its relationship with the motif occurrence probability may allow us to estimate a desired occurrence probability value and apply it as a target value to the motif-set selection process.

\subsection{Algorithmic Attacks}\label{subsec:Algorithmic Attacks}

It is claimed that the Aho-Corasick algorithm is less prone to algorithmic complexity attacks because of its deterministic performance \citeb{bremler2011space}. The same can be said of the Bouma2 Fast-Path algorithm, which acts as \quotes{the first line of defense} against such attacks. Nevertheless, while the choice of motifs can be optimized for a specific input, an attacker may generate malicious input containing \quotes{well-known} motifs that will trigger many false-positives. Solutions for this problem are being investigated, including throttling Slow-Path matches according to motif occurrence rates.

\subsection{Applying Statistics}\label{subsec:Applying Statistics}

Currently, occurrence statistics are applied only in the first preprocessing stage (see Section~\ref{subsec:Optimizing Motif Selection}). We believe it would be beneficial also to perform duplicates-removal and implement the \verb|B2-CALC-SCORING-OFFSET()| method in Algorithm~\ref{alg:B2-BUILD-SUBTRIE} using occurrence statistics, but obviously the benefits require further research. Having said that, tuning the Branch-and-Cut algorithm performance is essential: the time required to arrive at a solution is not deterministic, and sometimes exceeds 30 minutes with no solution\footnote{This happened on our 5th test; see Section~\ref{sec:Experimental Results}} for certain sets of statistics. We hope to find an optimization, given that the ILP we describe here seems far simpler than the kind of problems that the general-purpose Branch-and-Cut algorithm was designed to solve.

\subsection{Short Strings}\label{subsec:Short Strings}

Currently, Bouma2 supports strings 3 symbols long and above. Nevertheless, it has been claimed (see \citeb{dimopoulos2007memory}) that one of Aho-Corasick algorithm's appealing features is its ability to handle short strings well, as opposed to existing alternatives. One proposed solution for Bouma2 is to map single symbols to up to \textbf{$2 \cdot |\Sigma|$} motifs ($|\Sigma|$ motifs for even offsets and $|\Sigma|$ motifs for odd offsets), and similarly map 2-symbol strings to up to \textbf{$1 + |\Sigma|$} motifs (1 motif for even offsets and $|\Sigma|$ motifs for odd offsets, essentially expanding to 3-symbol strings). The effect on memory and performance of this enhancement should be examined.

\subsection{Bouma3 and Beyond}\label{subsec:Bouma3 and Beyond}

Setting the motif width to 2 symbols is very convenient from a technical point of view when dealing with 8-bit symbols ($|\Sigma|=256$), as in the case of Internet traffic inspection or file contents inspection. When we consider Computational Biology, the problem-space dictates $|\Sigma|=4$. Determining the motif width in this case should be done based on the trade-off between memory and performance constraints on one hand (wider motifs require more Fast-Path memory and would decrease Fast-Path performance because of cache-misses) and motif uniqueness on the other hand. Obviously, wider motifs require more than 2 mappings per word, and also more complicated short-string special case handling.

%% file: b2-appendix.tex
\appendix

\section{Mangled-Trie Construction Algorithm}\label{sec:Mangled-Trie Construction Algorithm}
Algorithms~\ref{alg:BOUMA2-BUILD-MANGLED-TRIE},~\ref{alg:B2-BUILD-SUBTRIE},~\ref{alg:B2-PURGE-OFFSET} and~\ref{alg:B2-FIND-PIVOT} detail the construction of a mangled-trie from a given resolve-set, as described in Section~\ref{subsec:Mangled-Trie Construction}. Note that the scoring offset calculation procedure and the actual mangled-trie construction procedures are not shown here because they are implementation-specific. We denote the symbol at offset $i$ relative to the motif position in a word $w$ as $\alpha(w, i)$.
%
%%%%%%%%%%%%%%%%%%%%%%%%%%%%%%%%%%%%%%%%%%%
\begin{algorithm}[h]
\DontPrintSemicolon
	\KwIn{$R_{\mu}$}
	\KwOut{$MT_{\mu}$}
	\Begin
	{
		\nlset{REM} Initialize with complete set of offset-symbol pairs:\;
		$S^{\mu} \leftarrow \Sigma^{MT_\mu}$\;
		$MT_{\mu} \leftarrow \text{B2-BUILD-SUBTRIE}(R_{\mu}, S^{\mu})$\;
		\nlset{REM} Optimize memory by removing duplicate nodes:\;
		$\text{B2-CONSOLIDATE-NODES}(MT_{\mu})$\;
	}
\caption{BOUMA2-BUILD-MANGLED-TRIE\label{alg:BOUMA2-BUILD-MANGLED-TRIE}}
\end{algorithm}
%%%%%%%%%%%%%%%%%%%%%%%%%%%%%%%%%%%%%%%%%%%%
%
%%%%%%%%%%%%%%%%%%%%%%%%%%%%%%%%%%%%%%%%%%%
\begin{algorithm}
\DontPrintSemicolon
	\KwIn{$R'_{\mu}, S$}
	\KwOut{$MT$}
	\Begin
	{
		$MT \leftarrow \emptyset$\;
		\If{$|R'_{\mu}| > 1$}
		{
			\nlset{REM} Heuristic for determining best offset for resolving:\;
			$i^{scoring} \leftarrow \text{B2-CALC-SCORING-OFFSET}(S)$\;
			$A \leftarrow \{\alpha(w, i^{scoring}): w \in R'_{\mu}\} \cup \{\epsilon\}$\;
			\ForEach{$\alpha^{consumed} \in A$}
			{
				$(W, S', w^{trans}) \leftarrow \text{B2-PURGE-OFFSET}(S, i^{scoring}, \alpha^{consumed})$\;
				\If{$w^{trans} \neq \epsilon$}
				{
					\nlset{REM} A transitional match occurred while consuming:\;
					$\text{B2-ADD-TRANSITIONAL}(MT, w^{trans})$\;
				}
				$(W^{pivot}, S^{pivot}) \leftarrow \text{B2-FIND-PIVOT}(S', i^{scoring})$\;
				\If{$|W^{pivot}| = 0$}
				{
					$MT' \leftarrow \text{B2-BUILD-SUBTRIE}(W, S')$\;
					$\text{B2-ADD-SUBTRIE}(MT, i^{scoring}, \alpha^{consumed}, MT')$\;
				}
				\Else
				{
					\nlset{REM} The Pivot recursively branches along 2 sides of motif:\;
					$MT' \leftarrow \text{B2-BUILD-SUBTRIE}(W \setminus W^{pivot}, S' \setminus S^{pivot})$\;
					$\text{B2-ADD-SUBTRIE}(MT, i^{scoring}, \alpha^{consumed}, MT')$\;
					$MT' \leftarrow \text{B2-BUILD-SUBTRIE}(W^{pivot}, S^{pivot})$\;
					$\text{B2-ADD-PIVOT-SUBTRIE}(MT, i^{scoring}, MT')$\;
				}
			}
		}
		\Else
		{
			$\text{B2-ADD-TERMINAL}(MT, W)$\;
		}
	}
\caption{B2-BUILD-SUBTRIE\label{alg:B2-BUILD-SUBTRIE}}
\end{algorithm}
%%%%%%%%%%%%%%%%%%%%%%%%%%%%%%%%%%%%%%%%%%%
%
%%%%%%%%%%%%%%%%%%%%%%%%%%%%%%%%%%%%%%%%%%%
\begin{algorithm}
\DontPrintSemicolon
	\KwIn{$S, i^{scoring}, \alpha^{consumed}$}
	\KwOut{$(W, S', w^{trans})$}
	\Begin
	{
		$w^{trans} \leftarrow \epsilon$\;
		\nlset{REM} All words with symbol mismatch at scoring offset:\;
		$\bar{W} \leftarrow \{w: \alpha(w, i^{scoring}) \notin \{\alpha^{consumed}, \epsilon\}\}$\;
		\nlset{REM} Inverse of $\bar{W}$:\;
		$W \leftarrow R'_{\mu} \setminus \bar{W}$\;
		\nlset{REM} Look for transitional:\;
		\If{$\exists w' \in W: \alpha(w', i') = \epsilon, \ \forall i' \neq i^{scoring}$}
		{
			$w^{trans} \leftarrow w'$\;
		}
		\nlset{REM} Discard all symbols at scoring offset:\;
		$S' \leftarrow S \setminus \{(w, i^{scoring}): w \in W\}$\;
		\nlset{REM} Discard all words with symbol mismatch at scoring offset:\;
		$S' \leftarrow S' \setminus \{(w, i): (w, i) \in S', w \in \bar{W}\}$\;
	}
\caption{B2-PURGE-OFFSET\label{alg:B2-PURGE-OFFSET}}
\end{algorithm}
%%%%%%%%%%%%%%%%%%%%%%%%%%%%%%%%%%%%%%%%%%%
%
%%%%%%%%%%%%%%%%%%%%%%%%%%%%%%%%%%%%%%%%%%%
\begin{algorithm}
\DontPrintSemicolon
	\KwIn{$S, i^{scoring}$}
	\KwOut{$(W^{pivot}, S^{pivot})$}
	\Begin
	{
		$S^{positive} \leftarrow \{(w, i): (w, i) \in S, i \geq 2\}$\;
		$W^{positive} \leftarrow \{w: (w, i) \in S^{positive}\}$\;
		$S^{negative} \leftarrow \{(w, i): (w, i) \in S, i \leq -1\}$\;
		$W^{negative} \leftarrow \{w: (w, i) \in S^{negative}\}$\;
		\If{$W^{positive} \cap W^{negative} = \emptyset$}
		{
			\If{$i^{scoring} \geq 2$}
			{
				$W^{pivot} \leftarrow W^{negative}$\;
				$S^{pivot} \leftarrow S^{negative}$\;
			}
			\Else
			{
				$W^{pivot} \leftarrow W^{positive}$\;
				$S^{pivot} \leftarrow S^{positive}$\;
			}
		}
	}
\caption{B2-FIND-PIVOT\label{alg:B2-FIND-PIVOT}}
\end{algorithm}
%%%%%%%%%%%%%%%%%%%%%%%%%%%%%%%%%%%%%%%%%%%
%

\section{Match-Process Algorithm}\label{sec:Match-Process Algorithm}
Algorithms~\ref{alg:B2-MATCH-PROC},~\ref{alg:B2-FAST-PATH},~\ref{alg:B2-SLOW-PATH} and~\ref{alg:B2-SP-LOOP} describe the Bouma2 match-process, as explained in Section~\ref{sec:The Match Process}. Implementation-specific procedures for matching symbols and substrings are omitted.
%
%%%%%%%%%%%%%%%%%%%%%%%%%%%%%%%%%%%%%%%%%%%
\begin{algorithm}[h]
\DontPrintSemicolon
	\KwIn{$M_L, \bigcup_{\mu \in M_L}\{MT_\mu\}, W_I \in \Sigma^*$}
	\KwOut{$\text{MATCHES}$}
	\Begin
	{
		$\text{HARVEST} \leftarrow \text{B2-FAST-PATH}(M_L, W_I)$\;
		$\text{MATCHES} \leftarrow \text{B2-SLOW-PATH}(\text{HARVEST}, \bigcup_{\mu \in M_L}\{MT_\mu\}, W_I)$\;
	}
\caption{B2-MATCH-PROC\label{alg:B2-MATCH-PROC}}
\end{algorithm}
%%%%%%%%%%%%%%%%%%%%%%%%%%%%%%%%%%%%%%%%%%%
%
%%%%%%%%%%%%%%%%%%%%%%%%%%%%%%%%%%%%%%%%%%%
\begin{algorithm}
\DontPrintSemicolon
	\KwIn{$M_L, W_I \in \Sigma^*$}
	\KwOut{$\text{HARVEST}$}
	\Begin
	{
		$\text{HARVEST} \leftarrow \emptyset$\;
		\ForEach{$i \in \{x:\ 0 \le x < |W_I|\ \land \ x \% 2 = 0\}$}
		{
			\If{$W_I = W_{I|p}tW_{I|s}:\ |W_{I|p}| = i\ \land \ t \in M_L$}
			{
				$\text{HARVEST} \leftarrow \text{HARVEST} \cup \{(i,t)\}$
			}
		}
	}
\caption{B2-FAST-PATH\label{alg:B2-FAST-PATH}}
\end{algorithm}
%%%%%%%%%%%%%%%%%%%%%%%%%%%%%%%%%%%%%%%%%%%
%
%%%%%%%%%%%%%%%%%%%%%%%%%%%%%%%%%%%%%%%%%%%
\begin{algorithm}
\DontPrintSemicolon
	\KwIn{$\text{HARVEST}, \bigcup_{\mu \in M_L}\{MT_\mu\}, W_I \in \Sigma^*$}
	\KwOut{$\text{MATCHES}$}
	\Begin
	{
		$\text{MATCHES} \leftarrow \emptyset$\;
		\ForEach{$(i,\mu) \in \text{HARVEST}$}
		{
			$MT_\mu \leftarrow \text{B2-MT-NEXT-TRANSITION}(MT_\mu, i)$\;
			$(\text{MATCHES}, MT_\mu^{pivot}) \leftarrow \text{B2-SP-LOOP}(MT_\mu, i, \text{MATCHES})$\;
			\If{$MT_\mu^{pivot} \neq \emptyset$}
			{
				$(\text{MATCHES}, MT_\mu^{pivot}) \leftarrow \text{B2-SP-LOOP}(MT_\mu^{pivot}, i, \text{MATCHES})$\;
			}
		}
	}
\caption{B2-SLOW-PATH\label{alg:B2-SLOW-PATH}}
\end{algorithm}
%%%%%%%%%%%%%%%%%%%%%%%%%%%%%%%%%%%%%%%%%%%
%
%%%%%%%%%%%%%%%%%%%%%%%%%%%%%%%%%%%%%%%%%%%
\begin{algorithm}
\DontPrintSemicolon
	\KwIn{$MT_\mu, i, \text{MATCHES}$}
	\KwOut{$(\text{MATCHES}, MT_\mu^{pivot})$}
	\Begin
	{
		\While{$MT_\mu \neq \emptyset$}
		{
			\If{$\text{B2-IS-TERMINAL}(MT_\mu)$}
			{
				$\text{MATCHES} \leftarrow \text{MATCHES} \cup \text{B2-MATCH-TERMINAL}(MT_\mu, i)$
			}
			\ElseIf{$\text{B2-IS-TRANSITIONAL}(MT_\mu)$}
			{
				$\text{MATCHES} \leftarrow \text{MATCHES} \cup \text{B2-MATCH-TRANSITIONAL}(MT_\mu, i)$
			}
			\ElseIf{$\text{B2-IS-PIVOT}(MT_\mu)$}
			{
				$MT_\mu^{pivot} \leftarrow \text{B2-GET-PIVOT}(MT_\mu)$\;
			}
			$MT_\mu \leftarrow \text{B2-MT-NEXT-TRANSITION}(MT_\mu, i)$\;
		}
	}
\caption{B2-SP-LOOP\label{alg:B2-SP-LOOP}}
\end{algorithm}
%%%%%%%%%%%%%%%%%%%%%%%%%%%%%%%%%%%%%%%%%%%
%